\documentclass[12pt,a4paper]{article}
\usepackage[textwidth=450pt, textheight=660pt]{geometry}
\usepackage{amsmath,amsthm,amssymb}
\usepackage{graphicx,tabu}
\usepackage[pagebackref=false]{hyperref}
\usepackage[active]{srcltx}
\usepackage[affil-it]{authblk}
\usepackage[noadjust]{cite}
\usepackage{filecontents}

 \newcommand{\be}{\begin{equation}}
\newcommand{\ee}{\end{equation}}
\newcommand{\bea}{\begin{eqnarray}}
\newcommand{\eea}{\end{eqnarray}}
\newcommand{\nn}{\nonumber}

\newcommand{\trm}{\textrm}
\newcommand{\tit}{\textit}
\newcommand{\mc}{\mathcal}

\newcommand{\ba}{\begin{array}}
\newcommand{\ea}{\end{array}}
\newcommand{\bfig}{\begin{figure}}
\newcommand{\efig}{\end{figure}}

\begin{document}

  \vspace{2cm}

  \begin{center}
    \font\titlerm=cmr10 scaled\magstep4
    \font\titlei=cmmi10 scaled\magstep4
    \font\titleis=cmmi7 scaled\magstep4
  {\bf Casimir temperature correction to the Bosonic string  mass }

    \vspace{1.5cm}
     \noindent{{\large Y. Koohsarian ${}^a$ \footnote{yo.koohsarian@mail.um.ac.ir},  A. Shirzad  ${}^{b,c}$ \footnote{shirzad@ipm.ir}}} \\
     ${}^a$ {\it Department of Physics, Ferdowsi University of Mashhad \\
       P.O.Box 1436, Mashhad, Iran} \\
   ${}^b$ {\it Department of Physics, Isfahan University of Technology \\
       P.O.Box 84156-83111, Isfahan, Iran,\\
      ${}^c$  School of Physics, Institute for Research in Fundamental Sciences (IPM),\\
       P. O. Box 19395-5531, Tehran, Iran} \\

  \end{center}
  \vskip 2em

\begin{abstract}
Through the Casimir effect, we obtain a temperature correction  to the ground state mass of the Bosonic string. Accordingly, we show that  higher than the temperature $\approx 1/2\pi $, the Tachyon attain real mass, and also we find  massive Graviton for nonzero temperatures. As another consequence, we demonstrate that for temperatures higher than the Hagedorn temperature, the string  would find again a Tachyonic state.
\end{abstract}


\section{Introduction} \label{sec-1}

Nowadays the Casimir effect is considered as one of the most direct  observable evidences for the fact that the \emph{vacuum } is not just  nothing, but rather it contains a sea of virtual entities known as the \emph{zero-point} oscillators of  quantum fields. In Casimir's seminal paper, the zero-point oscillations of the quantum electromagnetic field result in an attractive force between two parallel  chargeless conductor planes at zero temperature \cite{Cas}. Nowadays, the Casimir effect has been investigated  for different   systems  with  different boundary conditions, and different geometries, see e.g. \cite{CR1,CR2,CR3,CR4,CR5,CR6} as  reviews.  

In the framework of the Bosonic string theory, as we know, the quantum  zero-point oscillations of the string coordinates, result in  a negative energy for the ground state of the string. This negative zero-point energy gives the lowest mass state of the Bosonic string, i.e. the known Tachyon, an unusual  particle with imaginary mass. Besides its importance for describing the D-brane action in the string theory \cite{TachSt1,TachSt2,TachSt3,TachSt4,TachSt5}, it has been suggested that the Tachyon can play an important role in cosmology \cite{TachCos1,TachCos2,TachCos3,TachCos4,TachCos5}, specifically as a source of the dark energy  \cite{TachCos6,TachCos7,TachCos8,TachCos9,TachCos10}. However, finding an appropriate physical interpretation for the imaginary-mass  of the Tachyon is still a challenging problem for the   string theory \footnote{It has been shown that  the Tachyonic modes can also appear in some models of superstrings } . It has been conjectured that the existence of an open Bosonic  string Tachyon can be interpreted as the  instability and consequently the decay of a D25-brane \cite{Tach1,Tach2,Tach3}. However, the physical explanation of the closed string Tachyon has been even far less understood than the open string Tachyon \cite{Tach4} 

In this paper,  we tend  to introduce a new physical interpretation for the Tachyon, based on the Casimir effect. In fact we reinterpret the   negative mass-squared of the Bosonic string Tachyon, as the Casimir energy of the string coordinates. As a result, by finding the Casimir energy at nonzero temperatures, we would be able to obtain a   temperature correction to the Tachyon mass, which as we will see, this temperature correction has positive values for any nonzero temperature, and increases by increasing the temperature. Hence for sufficiently large temperatures, the negative mass-squared of  the Tachyon turns into a positive value, i.e. higher than a certain temperature the Tachyon  attains real mass, and thus the instable D25-brane would become stable.  But this temperature correction to the Tachyon mass, being taken as an  energy shift to the  string ground state, results in a nonzero mass for the Graviton, which is another interesting result, since the Graviton  is massless in the formal framework of the string theory, while massive gravitons have important roles for some Gravity theories, see e.g. \cite{MassG1,MassG2,MassG3}.  Finally  we investigate how these temperature corrections can be related  to the limiting temperature i.e. the known Hagedorn temperature of the Bosonic string.

\section{Tachyon mass as the Casimir energy }

We begin with the known expression of  the Bosonic string mass 
 \be
\alpha'  M^2=\int_0^{\pi} d\sigma \mathcal{H}(\dot{X},X')
\label{eq-1}
\ee
where  $\alpha'$ is the Regge slope parameter, $X^{\mu}(\sigma,\tau$)'s are the string coordinates  with $\sigma,\tau$ as the world-sheet parameters, and $\mc{H}$ represents the string Hamiltonian
\be
\mathcal{H}(\dot{X},X')=\frac{1}{4\pi \alpha'}  \left(\dot{X}^2+X'^2 \right).
\label{eq-2}
\ee
 Quantizing the string coordinates  in terms of the creation-annihilation operators of the harmonic oscillator ($a^{\dag}, a$), one  finds the familiar expression
\be
\alpha'  M^2=\sum_{I=1}^{d} \sum_{n=1}^{\infty} \left(a_n^{I\dag} a_n^I +\frac12 \right) n 
\label{eq-3}
\ee
 in which $d=25$ is the space dimension. The right-hand side  of the above equation, is obviously  a sum of Hamiltonians for an  infinite number of $24$-dimensional harmonic oscillators, with oscillation frequencies $\omega_n =n$. These harmonic oscillators are obviously the discrete harmonic-oscillation modes of the string (transverse) coordinates $X^I$'s .
  
  For the string ground state, one can simply find
 \be
\alpha'  M_0^2 = \int_0^{\pi} d\sigma \langle 0|\mathcal{H}|0 \rangle=\frac12 \sum_I \sum_n n 
\label{eq-4}
\ee
The divergent sum in  the above equation, is obviously the sum of zero-point energies of harmonic oscillation modes of the string coordinates, so one can take it as the zero-point energy of the string coordinates,
 \be
E_0=\frac12 \sum_I \sum_n n .
\label{eq-5}
\ee
 Now this divergent sum, can be regularized by several various  schemes, such as   Green function scheme \cite{CR3}, zeta function scheme \cite{math2}, Abel-Plana formula \cite{math1} scheme, etc. In the formal framework of the string theory, it is conventional  to regularize the above divergent sum, directly by applying the known (Riemann) zeta function,
\be
\zeta(s)=\sum_{n=1}^{\infty} n^{-s}.
 \label{eq-6}
\ee
which can be analytically continued for  arbitrary complex $s$, see e.g. \cite{math2}.  Particularly one can find
\be
\zeta(-2 l)=0, \ \  \zeta(1-2 l)=-\frac{B_{2l}}{2l}, \ \ l=1,2,...
\label{eq-7}
\ee
where $B$'s are the known Bernoulli numbers, which e.g.  we have  $B(2)=1/6$, $B(4)=-1/30$, etc.

  However, the divergent sum in Eq. \eqref{eq-5},  can also be regularized \emph{equivalently}  by using the  known  Abel-Plana formula: 
\be
\sum_{n=1}^{\infty} f(n)= \int_0^{\infty}f(k) dk +\frac{f(0)}{2} +i \int_0^{\infty}dk \ \frac{f(ik)- f(-ik)}{e^{2\pi k}-1} 
\label{eq-8}
\ee
with a function $f$ being analytic  in the right half-plane, see e.g \cite{math1}. The advantage of the Abel-Plana formula is that it  exhibits explicitly the infinite part of the divergent sum, i.e. just the first term in the right-hand side of  the above equation, while in the zeta function scheme, this infinite part would be removed automatically. Note that, taking $f(n)=n^{2l-1}$ in the Abel-Plana formula \eqref{eq-8},  one can show that
\be
\sum_{n=1}^{\infty} n^{2l-1}- \int_0^{\infty}k^{2l-1} dk=\zeta(1-2l) ; \ \ \ l=1,2,... 
\label{eq-9}
\ee
This equation can be verified directly for various values of $l$, which e.g. for $l=1$ one simply  finds
\be
\sum_{n=1}^{\infty} n - \int_0^{\infty}k  dk= -\frac{1}{12}=\zeta(-1).
\label{eq-10}
\ee
  In the framework  of the string theory, as we previously mentioned, the divergent sum in the string mass \eqref{eq-4} is regularized directly by using the zeta function  \eqref{eq-6}, to find the known mass formula of the Bosonic  string Tachyon
   \be
\alpha'  M_0^2  =  \sum_I \frac{\zeta(-1)}{2},
\label{eq-11}
\ee
  and so,  no physical explanation is given for the infinite part of the sum. Here we want  to show that this infinite part can be reasonably explained through the framework of the Casimir effect. In fact, as can be  comprehended from Eqs. \eqref{eq-2} and \eqref{eq-4}, the mentioned divergent sum can be taken naturally as the  zero-point energy of the quantum fields $X^{\mu}$'s, on which,   specific  boundary conditions imposed  at the boundary points $\sigma=0,\pi$ (for open string), or $\sigma=0,2\pi$ (for closed string). Now, the similarity to the conventional models of the  Casimir effect, would be more clear, if we change the world-sheet parametrization e.g. as
\be
\sigma \rightarrow \frac{\pi}{l_s} \sigma
\label{eq-12}
\ee
in which, $l_s$ is the string  characteristic length. Then Eq. \eqref{eq-4} turns to
  \be
 \int_0^{l_s} d\sigma \langle 0|\mathcal{H}|0 \rangle= \frac{\pi}{2 l_s} \sum_{I=1}^{24} \sum_{n=1}^{\infty} n
\label{eq-13}
\ee
This is just the zero-point energy of a massless 24-component vector field restricted to  the interval $0 <\sigma< l_s$  of a 1-dimensional $\sigma$-space, with a Hamiltonian just similar to Eq. \eqref{eq-2}, so we can write
  \be
E_0=\frac{\pi}{2 l_s} \sum_{I=1}^{24} \sum_{n=1}^{\infty} n
\label{eq-14}
\ee
 Then using Eq. \eqref{eq-8} one can write
\be
\frac{\pi}{2l_s}\sum_{n=1}^{\infty} n= \frac{\pi}{2l_s} \int_0^{\infty}k dk -\frac{\pi}{24l_s}
\label{eq-15}
\ee
This is a familiar equation in the framework of the Casimir effect. The second term in the  right-hand side of the above equation gives  just  the Casimir energy of the massless vector field,
\be
E_{\textrm{C}}=-\sum_I \frac{\pi}{24 l_s}=-\frac{\pi}{l_s}
 \label{eq-16}
\ee
Then, the first term, i.e. the  infinite integral term, can be interpreted, as is conventional in the framework of the Casimir effect, as the zero-point energy contribution of  the   field $X^{\mu}$ in the \tit{entire} 1-dimensional $\sigma$-space $(-\infty<\sigma<\infty)$,
\be
E_{0,\textrm{en}}=\sum_I \frac{\pi}{2 l_s} \int_0^{\infty}k dk
 \label{eq-17}
\ee
This is a reasonable interpretation, since in the entire $\sigma$-space, the discretized oscillation modes of the field $X^{\mu}$, would become continuous, turning the $n$-sum  to a $k$-integral, i.e. just the integral term in Eq. \eqref{eq-15}. Consequently one can write
\be
E_{\textrm{C}}=E_0-E_{0,\textrm{en}}
\label{eq-18}
\ee
Now, turning back to the formal parametrization  (of the world-sheet), one can obtain
\bea
&&E_0=\frac12\sum_I \sum_{n=1}^{\infty} n, \nn \\
&& E_{0,\textrm{en}}=\frac12 \sum_I \int_0^{\infty}k dk \nn \\
 &&E_{\textrm{C}}=\frac12 \sum_I \left(\sum_{n=1}^{\infty} n-\int_0^{\infty}k dk \right)
 \label{eq-19}
 \eea
  But using Eq. \eqref{eq-10}, the Casimir energy in the above equation can be written as
 \be
  E_{\textrm{C}}=\frac12 \sum_I \zeta(-1) 
  \label{eq-20}
  \ee
  which is just the right-hand side of Eq. \eqref{eq-11}. Hence the ground-state  mass formula of the open Bosonic string, can be written as
  \be
  \alpha' M_0^2 =  E_{\textrm{C}},
  \label{eq-21}
  \ee
 that is, the ground-state mass of the Bosonic string is given actually by the Casimir energy of the string coordinates $X^{\mu}$'s. The above interpretation is actually a physical explanation for removing the infinite part of the zero-point energy of the string. Making use of this interpretation, in the section 4, we obtain a temperature correction for the mass of the string. 

\section{String Casimir energy at finite temperature}

 To find the zero-point energy  at finite temperature, we use the known imaginary-time formalism, in which, the temperature is introduced  by the Matsubara frequencies, see e.g. \cite{CR2,CR3}. First, from the functional formalism of the quantum field theory, we can write the zero-point energy in  form \cite{IQFT,CR3}:
\be
E_0=\frac{i}{\textsf{T}}\ln{Z_0} 
\label{eq-22}
\ee
where $\textsf{T}$  is the total-time factor, and $Z_0$  represents the generating functional (in the absence of any source) which for the bosonic string  can be given, up to an irrelevant constant, as
\be
Z_0 \sim \prod_{I} \left[\det{\left( \partial^2_{\tau}- \partial^2_{\sigma} \right)}\right]^{-1/2}.  
\label{eq-23}
\ee
So the zero-point energy \eqref{eq-22}  takes the form
\be
E_0= -\frac{i}{2 \textsf{T}} \sum_{I}  \textrm{Tr} \left[\ln \left ( \partial^2_{\tau}- \partial^2_{\sigma}\right) \right], 
\label{eq-24}
\ee
where the trace `` $ \textrm{Tr}$'' is taken on the  modes
 \be
X_{\xi,n}(\sigma, \tau)=\sqrt{\alpha'} \exp(-i\xi \tau) \cos(n \sigma). 
\label{eq-25}
 \ee
Then, calculating the trace for the above mode, results in canceling the  total-time factor, and one would  find 
 \be
E_0= -\frac{i}{2} \sum_I \int_{-\infty}^{\infty} \frac{d\xi}{2\pi}  \sum_{n=1} ^{\infty} \ln\left[-\xi^2+ n^2 \right]. 
\label{eq-26}
 \ee
 To show that the  above equation is equivalent to the representation of  $E_0$ in  Eq. \eqref{eq-5}  we   rewrite the zero-point energy \eqref{eq-26} as
 \bea
E_0 &=&  \frac{ i}{2\pi} \sum_I  \sum_n \lim_{s \rightarrow 0} \frac{\partial }{ \partial s} \int_{0}^{\infty}d\xi \left(-\xi^2+ n^2 \right)^{-s}  \nn \\ &=& \frac{ i}{2\pi} \sum_I \lim_{s \rightarrow 0} \frac{\partial }{ \partial s} \zeta_R (2 s-1) \int_{0}^{\infty}dt \left(1-t^2 \right)^{-s},  \label{eq-27} 
 \eea
 where in the second line, we have introduced a new variable $\xi=t n$, and used  the  Riemann  zeta function \eqref{eq-6}. Then using
\bea
\int_{0}^{\infty}dt \left(1-t^2 \right)^{-s} = i\frac{\sqrt{\pi}}{2} \frac{\Gamma \left(s-\frac{1}{2}\right)}{\Gamma(s)}, 
\label{eq-28}
\eea
 see e.g. \cite{math4}, we find
 \bea
E_0= - \frac{1}{4\sqrt{\pi}} \sum_I  \lim_{s \rightarrow 0} \frac{\partial }{ \partial s} \frac{\Gamma \left(s-\frac{1}{2}\right)}{\Gamma(s)}\zeta_R (2 s-1), 
\label{eq-29}
 \eea
which equals  just  to the zero-point energy in  Eq.  \eqref{eq-5}. 

 Now, the temperature  can be introduced, as we previously mentioned, through the known imaginary-time formalism,  by imposing a   periodical condition for the imaginary-time parameter in the interval $[0,\beta]$, where $\beta=1/T$ with $T$ as the (single) string temperature. For the modes \eqref{eq-25}, which have been written in terms of  the real-time parameter $\tau$, this periodical condition can be written as
 \be
 X(\sigma,\tau-i\beta)=  X(\sigma,\tau).
 \label{eq-30}
 \ee
 Consequently one can find the   Matsubara (imaginary) frequencies 
\be
  -i\xi_m=2\pi m T \ \ \ ; \ \ \ m=0,\pm1,\pm2,... 
  \label{eq-31}
\ee
Thus the zero-point energy \eqref{eq-26} at finite temperature, takes the form
 \be
E_0(T) = \frac{T}{2}\sum_I\sum_{m=-\infty}^{\infty} \sum_{n=1} ^{\infty} \ln\left[(2\pi T)^2 m^2+ n^2 \right]. 
\label{eq-32}
 \ee
 So, similarly as Eq. \eqref{eq-18}, the finite-temperature Casimir energy of the open bosonic string, can be given as
 \bea
&& E_{\textrm{C}}(T) = E_0(T)-E_ {0,\textrm{en}}(T) \nn \\
 &&E_ {0,\textrm{en}}(T)  = \frac{T}{2}\sum_I\sum_{m=-\infty}^{\infty} \int_0^{\infty} dk \ln\left[(2\pi T)^2 m^2+ k^2 \right]. \label{eq-33}
 \eea
To find an explicit expression for $E_C(T)$, first we  rewrite the zero-point energy \eqref{eq-32} as  a parametric integral,
 \bea
E_0(T)&=&  -\frac{T}{2} \sum_I  \sum_{m=-\infty}^{\infty} \sum_{n=1}^{\infty} \lim_{s \rightarrow 0} \frac{\partial }{ \partial s}\left[(2\pi T)^2 m^2+ n^2 \right]^{-s} \nn \\
&=& -\frac{T}{2} \sum_I \lim_{s \rightarrow 0} \frac{\partial }{ \partial s} \int_0^{\infty}\frac{dt}{t} \frac{t^s}{\Gamma(s)} \sum_{n=1}^{\infty} \exp\left[-t n^2 \right] \nn \\
&&\hspace{3cm} \times\left( 1+2\sum_{m=1}^{\infty}\exp \left[ -t \left(2\pi T m\right)^2\right] \right) 
\label{eq-34}
 \eea
where in the third line we have extracted the term $m=0$ from the whole $m$-sum. Now, we apply the Poisson summation formula (see e.g. \cite{math3})
 \be
\sum_{n=1}^{\infty} f(n)= -\frac{f(0)}{2} + \int_0^{\infty} f(x) dx +2 \sum_{n=1}^{\infty} \int_0^{\infty} f(x) \cos(2\pi nx)dx \label{eq-35}
\ee
  to the $n$-sum, to obtain
  \be
\sum_{n=1}^{\infty}\exp \left[ -t  n^2\right]=-\frac{1}{2}+ \frac12 \sqrt{\frac{\pi}{t}}+\sqrt{\frac{\pi}{t}} \sum_{n=1}^{\infty} \exp{\left[-\frac{\pi^2 n^2}{ t} \right]}. 
\label{eq-36}
\ee
Then by substituting  this expression into Eq. \eqref{eq-34},    and after some calculations, one can find
 \bea
E_0(T)=-12 T  \lim_{s \rightarrow 0} \frac{\partial }{ \partial s} \Big[ \zeta (2s) -\frac{ \zeta (2s)}{(2\pi T)^{2s}}+ \sqrt{\pi} \frac{ \zeta (2s-1)}{(2\pi T)^{2s-1}}\frac{\Gamma(s-1/2)}{\Gamma(s)}+ \nn \\
 4\sqrt{\pi}\sum_{n,m=1}^{\infty}\left(2Tm/n\right)^{-s+1/2} \frac{K_{-s+1/2}(4 \pi^2 n m T)}{\Gamma(s)} \Big],
  \label{eq-37}
 \eea
in which, $K$ is a Bessel function of the second kind, and $\zeta $ is the Riemann zeta function \eqref{eq-6}, and we have utilized the integral relation
\be
\int_0^{\infty}  t^r \exp\left[-x^2 t-y^2/t \right]dt=2(x/y)^{-r-1}K_{-r-1}(2xy). 
\label{eq-38}
\ee
 Finally, taking the limit, we obtain
\be
E_0(T)=-4 \pi^2 T^2 +12 T \ln(2\pi T)- 24 T\sum_{n,l=1} \frac{\exp(- 4 \pi^2  n l T)}{n}
 \label{eq-39}
\ee
In a similar way, we can find
\be
E_{0,\textrm{en}}(T)=-4\pi^2 T^2 
\label{eq-40}
\ee
Thus the Casimir energy of the string at finite temperature would be given as
\be
E_C(T)=12T\ln{ 2\pi T}-24T \sum_{n,m=1}^{\infty} \frac{\exp(-4\pi^2 n m T)}{n}
 \label{eq-41}
\ee
Note that the above equation is an exact expression for the finite-temperature Casimir energy of the string, however, it can also provide an asymptotically explicit expression for the string Casimir energy at    high temperature limit:
\be
E_C(T) \approx 12T\ln{ T}  \ \  ; \ \ T \gg 1 .
\label{eq-42}
\ee
The above useful technique for obtaining an asymptotic expression for the  Casimir energy, is less complicated in comparison to the familiar heat-kernel approach (see e.g. \cite{math5}), and can be directly generalized  for many different configurations.

 To find an expression being asymptotically useful  for  low temperatures, we need to apply the Poisson formula  to the $m$-sum in Eq. \eqref{eq-34}, and after similar calculations, one can find
\be
E_C(T)=-1+4\pi^2 T^2-\sum_{n,m=1} \frac{\exp(-n m/T)}{n} 
 \label{eq-43}
\ee
which for sufficiently low temperatures can be approximated as
\be
 E_C(T) \approx -1+4\pi^2 T^2  \ \  ; \ \ T \ll 1 \label{eq-44}
\ee
Note that at zero temperature, the above equation gives $E_{\trm{C}}(0)=-1$ which is in agreement with Eq. \eqref{eq-20}, as expected.
\section{Casimir temperature correction to the  string mass}

\begin{figure}[b]
\centering
\includegraphics [width=12 cm, height=6  cm]{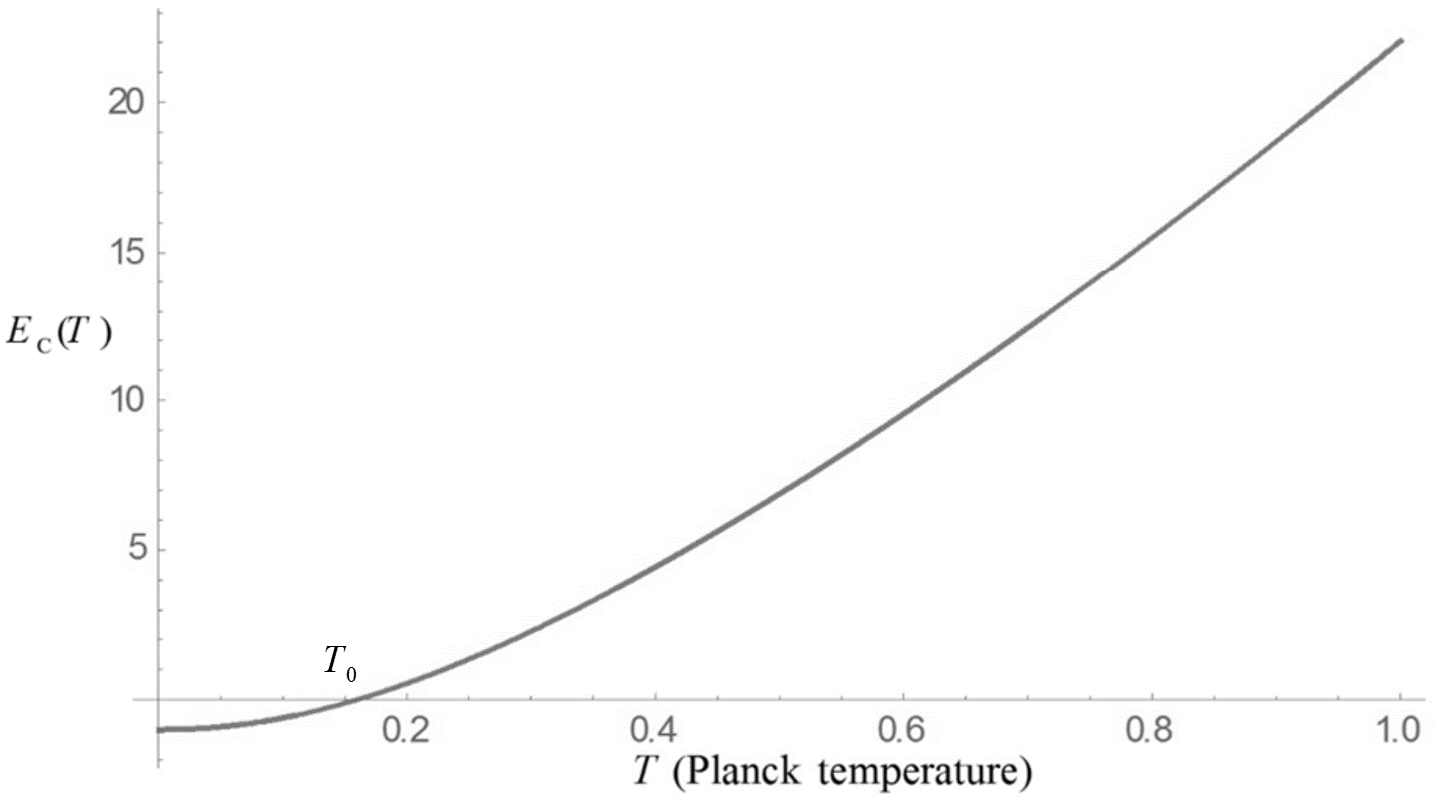}
\label{SCET}
\caption{String Casimir energy in terms of the temperature}
\end{figure}

As we know, the ground-state mass of the string is given by its zero-point energy. From the standpoint of the Casimir effect, as we  discussed previously  in Sec. 1, this zero-point energy can be interpreted as the  Casimir energy of the  string  coordinates. So for the open Bosonic string, one can write
\begin{equation}
\alpha' M_0^2 = E_C= \sum_{I=1}^{24} \frac{\zeta_R(-1)}{2}=-1,
 \label{eq-45}
\end{equation}
As a result,   we can generalize the above equation to find the ground-state mass of the  string at finite temperature
\be
\alpha' M_0^2(T) = E_C(T) ,
\label{eq-46}
\ee
with $E_C(T)$ given by either Eq. \eqref{eq-41} or Eq. \eqref{eq-43}.  Note that, from Eq. \eqref{eq-43}, we obviously have $\alpha' M_0^2(0)=-1$,  as expected. Therefore Eq. \eqref{eq-43}  provide a temperature  correction to the Tachyon mass of  the string. As is seen in the numerical  plot \ref{SCET}, higher than a certain (small) temperature $T_0$, $E_C(T)$ reaches the positive values, i.e.  the Tachyon attains real mass. But for sufficiently low temperatures, Eq. \eqref{eq-46} can be approximated  as
\be
\alpha' M_0^2(T) \approx -1+4 \pi^2 T^2 ,
 \label{eq-47}
\ee
see Eq. \eqref{eq-43}. Therefore we obviously find
\be
M_0^2 \geq 0 \ \ \  \textrm{for} \ \  T\geq T_0 \approx \frac{1}{2\pi},
 \label{eq-48}
\ee
that is, for temperatures higher than $T_0 \approx 1/2\pi $, the Tachyon would turn into a real particle, with mass given by Eq. \eqref{eq-46}. 

On the other hand, the temperature correction terms in Eq. \eqref{eq-46} provide actually a shift in the ground state energy of the string, thus the mass formula of the open Bosonic string, at nonzero temperature, can be taken as
\be
\alpha' M^2(T) =E_C(T) +N , \label{eq-49}
\ee
with $N=0,1,2,...$ .  Physically Eq. \eqref{eq-49}  can be   interpreted  as to say,  by increasing the temperature, part of  the thermal energy can  be consumed by the ground state of the string. 

 In a similar way, for a closed Bosonic string, we find
\be
\frac{\alpha'}{4} M^2(T) = E_C(T) +N  .
\label{eq-50}
\ee
Obviously, just as for  the open string, the closed string Tachyon  for temperatures higher than $ T_0 \approx 1/2\pi$ , would attain real mass, by $\alpha' M_0^2(T) =4 E_C(T) $. Subsequently one can find a temperature correction to the mass of Graviton ($N=1$)
\be
\frac{ \alpha'}{4} M_1^2(T) =  E_C(T) +1. \label{eq-51}
\ee
From Eq. \eqref{eq-43}, obviously $E_C > -1$ for $T>0$, thus at nonzero temperature, the Graviton attains mass by Eq. \eqref{eq-51}.

\section{Limiting temperature of the string}

The temperature can also be introduced, as we know, through the statistical mechanics by using the entropy function. As we know, at high energy limit, the  entropy of a single open Bosonic  string can be obtained, using the Hardy-Ramanujan  asymptotic expansion, as
\bea
S \approx 4\pi \sqrt{N} \ \  \ ; \ \ \ \textrm{for large value of $N$}, \label{eq-52}
\eea
in which $N$ is  the string  state number. In the rest frame of the open string,  Eqs.\ \eqref{eq-49} and \eqref{eq-52} result in
\bea
S (E,T) \approx 4\pi \left[\alpha' E^2 - E_C(T) \right]^{1/2} \ \  \ ; \ \ \ \textrm{for large values of $E$ and $T$} ,
\label{eq-53}
\eea
with $E$ as the string energy. At hight temperature, using Eq. \eqref{eq-42}, the entropy \eqref{eq-53} can be written as
\be
S (E,T) \approx 4\pi \left[\alpha' E^2 - 12 T \ln T\right]^{1/2} .
 \label{eq-54}
\ee
Now for the string temperature, we can write
\bea
\frac{1}{T} &=& \frac{\partial S}{\partial E} \nn \\
& \approx& 4 \pi \alpha' E \left[\alpha' E^2 - 12 T \ln T \right]^{-1/2}  .
\label{eq-55}
\eea
This equation can be solved for $E$,
\be
E^2 \approx \frac{12 T \ln T}{\alpha' -(4\pi \alpha' T)^2},  \label{eq-56}
\ee
then considering the denominator,  one can find a limiting temperature just equal to the Hagedorn temperature
\be
1 \ll T < \frac{1}{4\pi \sqrt{\alpha'}}= T_H .\label{eq-57}
\ee
Note  that $E^2<0$ for $T>T_H$, i.e. for temperatures higher than the Hagedorn temperature, the string  would find again a Tachyonic state. One can show that, the above result also holds for a single  closed string. Finally we can compare $T_0$ with  $T_{\textrm{H}}$ as  
\be
T_0 < 1 \ll T_{\textrm{H}}, \label{eq-58}
\ee
see Eqs. \eqref{eq-48} and \eqref{eq-57}. Thus the temperature $T_0$,  above which the Tachyon would attain real mass, is far below the Hagedorn temperature, and so is accessible. Note also that Eq. \eqref{eq-58}  impose actually a physical constraint on the Regge slope parameter.

\section{Conclusion and  remarks}
The main idea of this work is a new reinterpretation of the Tachyon mass through the Casimir effect. This interpretation leads us to some new interesting results :

 We have found a temperature correction to the mass of the open as well as closed Bosonic string, which has a  positive value for any  temperature, and which increases by increasing the temperature. We have shown that for temperatures higher than $ \approx1/2\pi$ Planck temperature, the  negative mass-squared of the  Tachyon would turns to positive values, i.e. the instable D25-brane would become stable.  We have obtained a temperature correction to the Graviton mass, so that for nonzero temperatures the Graviton would become massive, which is another  interesting result, since the Graviton  is massless in the formal framework of the string theory, while massive gravitons play important roles in some Gravity theories. We have found an asymptotic relation for the Bosonic string mass at high temperature, and have shown that for temperatures higher than the Hagedorn temperature, the string would find again a Tachyonic state. We also obtained a physical constraint on the Regge slope parameter.

\subsection*{Acknowledgments}
We thank Ahmad Ghodsi for his valuable comments, Amin Faraji Astaneh and Hessamaddin Arfaei for their useful comments,  and  Mohammad Moghadassi for his useful discussions.

%%%%%%%%%%%%%%%%%%%%%%%%%%%%%%%%%%%%%%%%%%%%%%%%%%%%%%%%%%%%%%%%%%%%%%%
\end{document}